\documentclass[onecolumn,preprint,amsmath,amssymb]{revtex4}

\usepackage{graphicx}
\usepackage{dcolumn}
\usepackage{bm}

\begin{document}

\newcommand{\lia}{\mathbf{l}_i^{(\alpha)}}
\newcommand{\Lia}{L_i^{(\alpha)}}
\newcommand{\Lib}{L_i^{(1)}}
\newcommand{\Lic}{L_i^{(2)}}
\newcommand{\mi}{\mathbf{m}_i}
\newcommand{\bn}{\mathbf{n}}
\newcommand{\bs}{\mathbf{s}}
\newcommand{\br}{\mathbf{r}}
\newcommand{\lb}{\ell^{(1)}}
\newcommand{\lc}{\ell^{(2)}}


\title{Instantaneous Liquid Interfaces}

\author{Adam P. Willard}
 \author{David Chandler}
\affiliation{Department of Chemistry, University of California, Berkeley, California 94720}

\date{\today}

\begin{abstract}
We describe and illustrate a simple procedure for identifying a liquid interface from atomic coordinates.  In particular, a coarse grained density field is constructed, and the interface is defined as a constant density surface for this coarse grained field.  In applications to a molecular dynamics simulation of liquid water, it is shown that this procedure provides instructive and useful pictures of liquid-vapor interfaces and of liquid-protein interfaces.
 \end{abstract}

\maketitle

\section*{The Interface}
Definitions of soft-matter interfaces at a molecular level can be ambiguous~\cite{EC03,DH87}.  Due to molecular motions, interfacial configurations change with time, and the identity of molecules that lie at the interface also change with time.  Generally useful procedures for identifying interfaces must accommodate these motions.  Here, we present a simple and intuitive procedure for doing so.   The procedure is based upon spatial coarse graining, it applies to reasonably arbitrary geometries, and it can be applied at any point in time so that it can be used to interpret time dependent phenomena and fluctuations.  We find the procedure to be useful in a variety of contexts, a few of which are illustrated in this and the next section.

The basic idea begins with the instantaneous density field at space-time point $\mathbf{r},t$, 
\begin{equation}
\label{density}
\rho (\mathbf{r},t)=\sum_i \delta \left( \mathbf{r} - \mathbf{r}_i(t)\right),
\end{equation}
where $\mathbf{r}_i(t)$ is the position of the $i$th particle at time $t$, and the sum is over all such particles of interest.  Rendering this field directly provides only vague impressions of interfaces.  A more manageable field can be formed through coarse graining.  Our choice of spatial coarse graining is a convolution with the normalized Gaussian functions
\begin{equation}
\label{gaussian}
\phi(\mathbf{r}; \xi) = (2\pi \xi^2)^{-d/2} \exp(-r^2 / 2 \xi^2),
\end{equation} 
where $r$ is the magnitude of $\mathbf{r}$, $\xi$ is the coarse graining length, and $d$ stands for dimensionality.  Applied to $\rho (\mathbf{r},t)$ we have the coarse grained density field
\begin{equation}
\label{coarse}
\bar{\rho}(\mathbf{r},t) = \sum_i \phi \left(|\mathbf{r} - \mathbf{r}_i (t)|; \xi \right).
\end{equation}
The choice of $\xi$ will depend upon the physical conditions under considerations.  With $\xi$ set, we define interfaces to be the $(d-1)$-dimensional manifold $\mathbf{r} = \mathbf{s}$ for which 
\begin{equation}
\label{interface}
\bar{\rho}(\mathbf{s}, t) = c,
\end{equation}
where $c$ is a constant.  In other words, we define instantaneous interfaces to be points in space where the coarse grained density field has the value $c$.  This coarse grained density changes with time as molecular configurations change with time, i.e., $\bs = \bs(t) =\bs(\{\br_i(t)\})$. 

\begin{figure}
\includegraphics[width=3.416in]{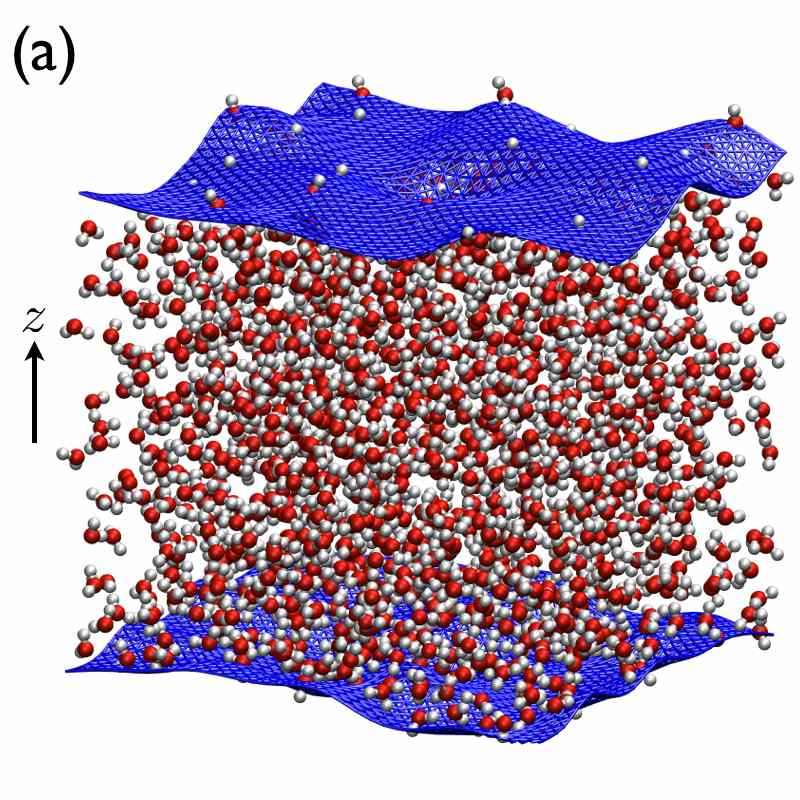}
\includegraphics[width=3.416in]{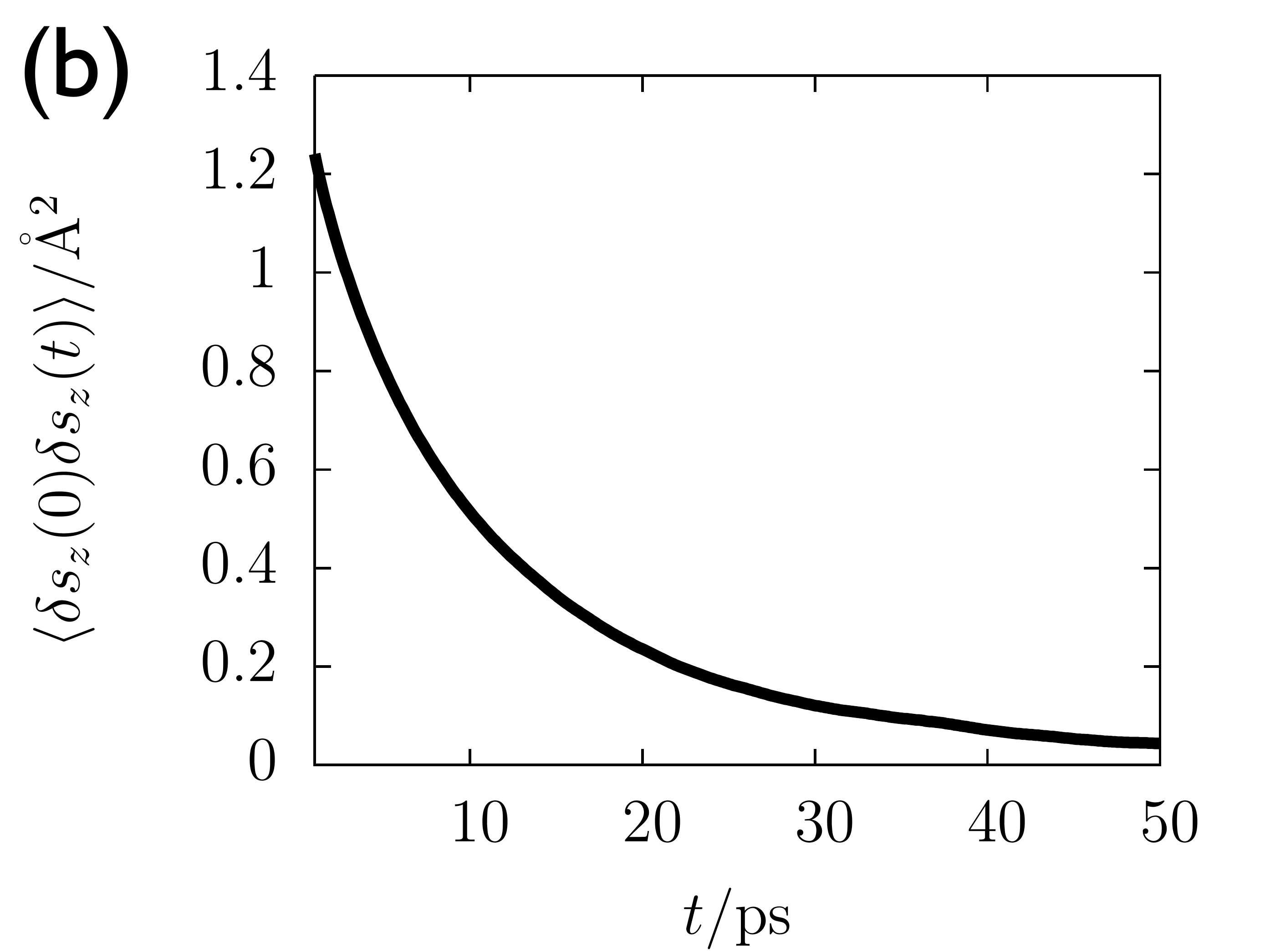}
\caption{(a)A snapshot of a slab of liquid water with the instantaneous interface $\bs$ rendered as a blue mesh on the upper and lower phase boundary. The slab is periodically replicated in the horizontal directions. (b)The time correlation function governing the spatial fluctuations in the intrinsic interface $\bs$. Here angle brackets represent an equilibrium average and $\delta s_z(t) \equiv (\bs(t) \cdot \hat{z} - \langle \bs \rangle \cdot \hat{z})$ where $\bs(t)$ is the position of the interface at time $t$ and $\hat{z}$ is the unit vector in the $z$ direction (as is indicated in Panel (a)). 
}
\label{fig:planar}
\end{figure}

\begin{figure}
\includegraphics[width=5.5in]{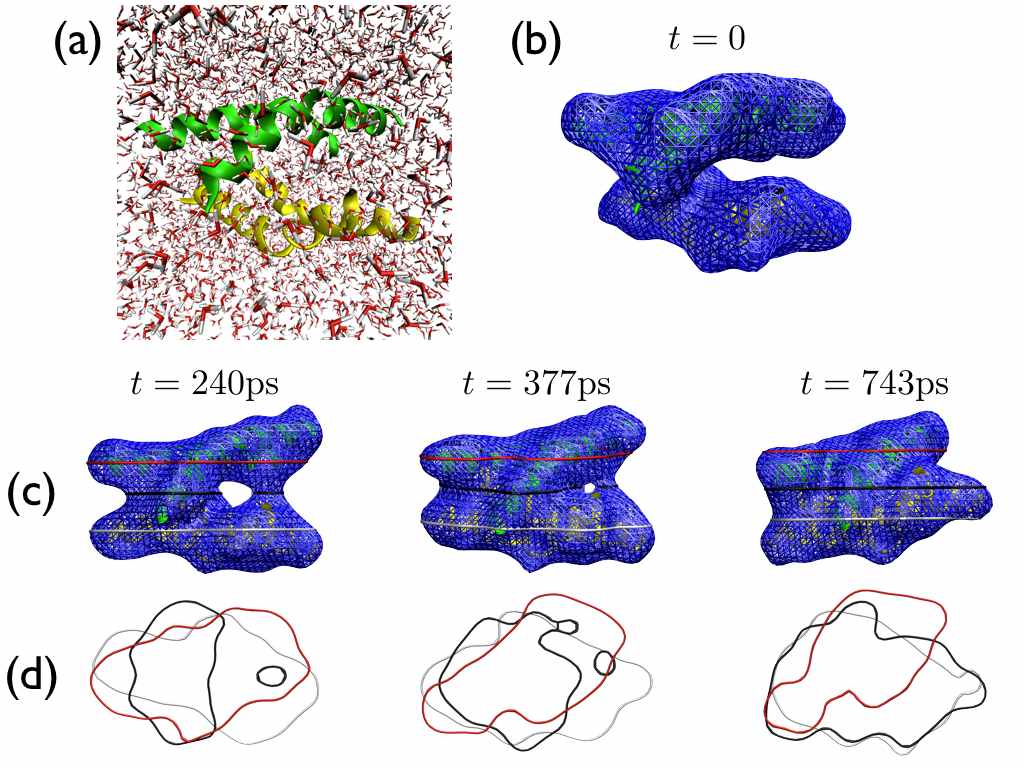}
\caption{Snapshots of an assembling pair of melittin dimers. (a)Snapshot at $t=0$ showing explicit solvent and melittin dimers rendered as green and yellow ribbons. (b)Snapshot at $t=0$ excluding the explicit solvent with the instantaneous interface $\bs$ rendered as a blue mesh. (c)Snapshots of interface at $t = 240 \mathrm{ps}$, $t = 377 \mathrm{ps}$, and $t = 743 \mathrm{ps}$. Red, blue, and white bands illustrate cross-sections in $\bs$ as seen from the top in (d).
}
\label{fig:melittin}
\end{figure}

For a given molecular configuration, $\{ \br_i(t) \}$, Eq.~(\ref{interface}) can be solved quickly through interpolation on a spatial grid. \ref{fig:planar}  illustrates what is found for one configuration of a slab of liquid water at conditions of water-vapor coexistence.  Details of our simulations are described below in the Methods section.  We have taken $\{ {\mathbf{r}_i(t)}\}$ to refer to the positions of all the oxygen atoms in the system, and because the bulk correlation length of liquid water is about one molecular diameter, we have used use $\xi = 2.4 $\AA; further, we have used $c = 0.016$\AA$^{-3}$, which is approximately one-half the bulk density. 

The pictured instantaneous water-vapor interface $\bs$ resembles a wavy sheet dividing the liquid and vapor-like regions. In accordance with the predictions of capillary wave theory~\cite{FB65, JW77, Widom}, the undulating height fluctuations in $\bs$ exhibit transverse correlations that extend across the system.  As shown in \ref{fig:planar}, for the size system pictured, these fluctuations de-correlate on time-scales of tens of picoseconds.

This approach belongs to a broader class of interface identification algorithms that build upon the assumption that the interface is well described as a molecularly sharp $d-1$ dimensional manifold that is made rough by collective thermal fluctuations~\cite{DH87,JW77,EC06,JF07,FU09,LP08,EC03}.  Our procedure is distinguished by being independent of an interfacial reference plane or presumed symmetry.  This feature is particularly useful for studying liquid interfaces near non-planar substrates and irregularly shaped solutes like biopolymers.  We turn to such an application in \ref{fig:melittin}, which shows the instantaneous interfaces of water during a trajectory of two hydrated melittin dimers.   This example has no obvious spatial symmetries.  

Melittin dimers have hydrophobic domains that are exposed to solvent until they assemble to form a stable tetramer.  In this assembly, the hydrophobic domains undergo a dewetting induced hydrophobic collapse~\cite{LP05,LH07,NG08}. The rendered interfaces in \ref{fig:melittin} show how this aggregation is highly collective.  The concerted motions of water molecules that underlie what is pictured would be hard to detect without viewing the instantaneous interfaces.  In the specific trajectory illustrated, the dimers first come into contact on one end. Collapse proceeds through a zipper-like motion during which water is squeezed out of the cavity away from the ends of the dimers that have already associated. This process is aided by the intermittent formation of vapor-tunnels bridging the unassociated ends of the dimers (examples of these vapor tunnels are shown at $t=240\mathrm{ps}$ and $t = 377 \mathrm{ps}$ in \ref{fig:melittin}).  Due to these vapor tunnels, there is an unbalancing of solvent-induced forces, and this unbalancing accelerates assembly~\cite{LCW, DC_Nature05, APW08}.

\section*{Contrasting Mean and Instantaneous Interfaces}
To quantify molecular properties associated with interfaces, it is useful to carry out averages in terms of the positions and orientations of molecules with respect to the location of the interface.  In such considerations, the distinction between mean and instantaneous interfaces is significant. This is because interfacial fluctuations can be large, so that a molecule located at the position of the mean interface can be often distant from the instantaneous interface.  Using the mean density or mean interfacial profile to specify the reference surface will therefore obscure the true molecular nature of interfacial properties.  To illustrate the significance, we have examined a few properties associated with the liquid-vapor interface.  In this case, the mean interface is essentially the fixed reference frame of the Gibbs dividing surface~\cite{Gibbs}.

In either frame of reference, we let $a_i(t)$ denote the proximity of the $i$th water molecule to the surface.  That is,  
\begin{equation}
a_i(t) = \{ \left[ \bs(t) - \br_i(t) \right] \cdot \bn(t)\} \vert_{\bs(t) = \bs_i^*(t)} ,\,\,\,\,\mathrm{instantaneous},
\end{equation}
or
\begin{equation}
a_i(t) = \left[ \langle \bs \rangle - \br_i(t) \right] \cdot \langle \bn \rangle,\,\,\,\,\mathrm{mean}.
\end{equation} 
The angle brackets denote equilibrium average; $\bs_i^*(t)$ is the point on $\bs(t)$ nearest $\br_i(t)$; $\bn(t)$ is the unit vector normal to the instantaneous interface at $\bs(t)$, i.e., the unit vector in the direction of $\nabla \bar \rho (\br, t) \vert_{\br = \bs(t)}$; $\langle \bn \rangle$ is the unit vector normal to the mean surface, $\langle \bs \rangle$; $a_i(t)$ is positive if molecule $i$ is on the vapor side of the interface and negative if molecule $i$ is on the liquid side of the interface.  [While we are here considering specifically liquid-gas coexistence, where the mean interface has planar symmetry, the proximity $a_i(t)$ defined in Eq. (5) is more generally applicable.]

\begin{figure}
\includegraphics[width=3.416in]{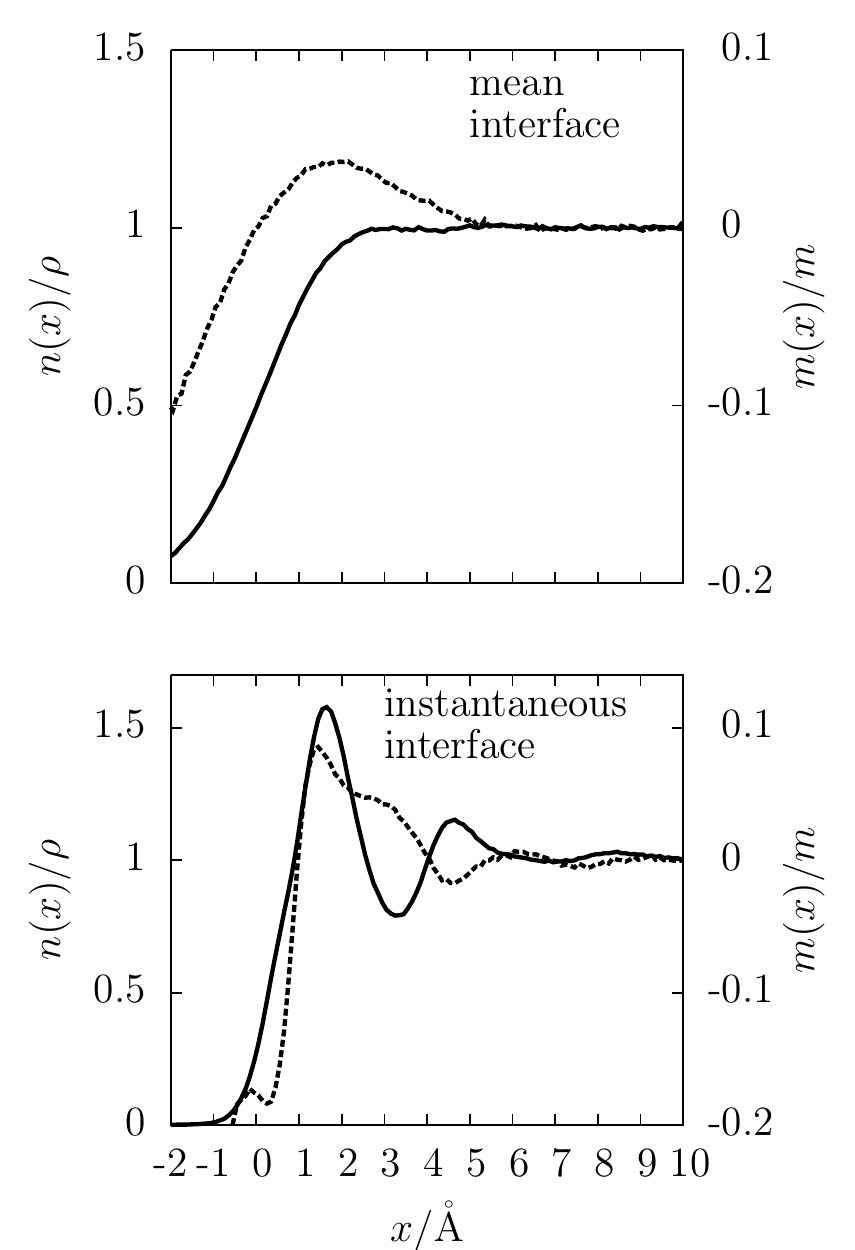}
\caption{The mean density profile $n(x)$(solid line) and the mean orientational property $m(x)$(dashed line) plotted as functions of the distance $x$ relative to either the Gibbs dividing surface (top) or the instantaneous interface (bottom). The quantity $\rho$ is the value of bulk liquid density and $m$ is the magnitude of the dipole moment of a single water molecule in the simulation.
}
\label{fig:denprof}
\end{figure}

With this notation, the mean density profiles with respect to either the instantaneous water-vapor interface or the mean water-vapor interface is
\begin{equation}
n(x) = \frac{1}{L^2} \left \langle \sum_{i}\delta ( a_i - x) \right \rangle,
\end{equation} 
where $L$ is the length of the simulation cell parallel to the mean interface. \ref{fig:denprof} juxtaposes the density profile when $a_i$ is in reference to the instantaneous interface, Eq. (5), with that when $a_i$ is in reference to the mean interface, Eq. (6).  The profile relative to the fluctuating instantaneous interface exhibits oscillations indicative of a layering of the atomistic solvent. 
This layering while not without precedent~\cite{FB08,EC06,PT02,EC01} represents a significant departure from the more familiar sigmoidal density profile observed relative to the mean dividing surface. It is evident from this result that the liquid-vapor phase boundary is indeed well described as a sharp surface (with a width of approximately a single molecular diameter) dividing the bulk liquid and vapor phases. In this perspective, proposed by Stillinger~\cite{FB65} and later treated extensively by Widom~\cite{BW69} and Weeks~\cite{JW77}, the sigmoidal feature seen in the mean solvent density profile (\ref{fig:denprof}) is a manifestation of thermal fluctuations in the position of the intrinsic surface. An important consequence arises in the context of extended hydrophobic interfaces whose density profiles are often drawn to resemble those of a liquid-vapor interface, meaning they including a region of depleted solvent density between the hydrophobic surface and the bulk liquid. It has been shown that for extended hydrophobic solutes that interact with water through even weak dispersive interactions that the solvent density profile is often not sigmoidal but exhibits layering~\cite{MP07,RN07,NC08,GH08,JW09,SG09,MP01}. This observation has been used to conclude~\cite{MP07,NC08,MP01} that water near such hydrophobic surfaces is not like water near a liquid-vapor interface. \ref{fig:denprof} shows that such a conclusion is not justified. In particular, the mean water density proximal to the instantaneous liquid-vapor interface is layered. The layering thus seen in the mean density near hydrophobic surfaces indicates that dispersive attractions are sufficient to locate the water-vapor interface on average adjacent to the surface. Without those weak attractions the interface would wander.

Not only does the layering indicate that the instantaneous $\bs$ neatly divides the vapor and liquid phases, the modulation in the solvent density has a significant effect on the orientations of water molecules in the vicinity of the water-vapor interface.  This orientational structure is washed out when the Gibbs or mean surface is used as the reference.  Consider, for instance the dipole moment of the $i$th water molecule, $\mathbf{m}_i$, for which the conditional mean of its projection parallel to the instantaneous or mean interface is given by
\begin{equation}
m(x) = \frac{1}{L^2 n(x)}\left \langle \sum_{i=1}^N m_i\, \delta ( a_i - x) \right \rangle.
\end{equation}
where either
\begin{equation}
m_i = \left( \mathbf{m}_i \cdot \bn \right)\vert_{\bs=\bs_i^*}, \,\,\,\,\mathrm{instantaneous}
\end{equation} 
or
\begin{equation}
m_i = \mathbf{m}_i \cdot \langle \bn \rangle , \,\,\,\,\mathrm{mean}.
\end{equation}
These quantities are plotted alongside their respective density profiles in \ref{fig:denprof}.  In the frame of reference of the instantaneous interface, oscillations of $m(x)$ with respect to $x$ reflects the correlations between molecular orientation and proximity to the interface.  These correlations are not nearly so evident or interpretable in the frame of reference of the mean interface. 

\begin{figure}
\includegraphics[width=6.8in]{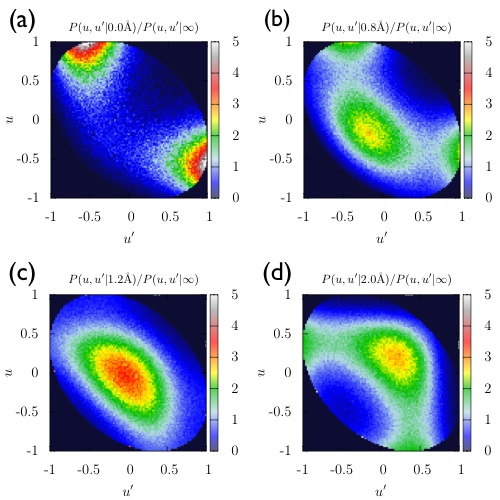}
\caption{The joint probability distribution $P(u,u' \vert x)$ governing the orientations of each OH bond of water molecules with $x=0$(a), $x=0.8$\AA(b), $x=1.2$\AA(c), and $x = 2.0 $\AA(d). A value of $\ell = 1$, $\ell = 0$, or $\ell = -1$ refer to OH bonds that are oriented parallel, perpendicular, or anti-parallel to the local surface normal respectively. The normalizing factor $P(u,u' \vert \infty)$ is the value of the joint probability distribution in an isotropic environment.
}
\label{fig:joint_prob}
\end{figure}

Indeed, orientational structure is sufficiently vivid in the reference frame of the instantaneous interface that we find it informative to also consider quantities like the joint conditional distribution
\begin{equation}
P(u,u'|x) \propto \frac{1}{n(x)} \left \langle \sum_{i} \delta \left( u - \cos (\theta_i^{(1)})\right) \delta \left( u' - \cos (\theta_i^{(2)}) \right) \delta \left(a_i - x \right) \right \rangle,
\end{equation}
where $\theta_i^{(1)}$ is the angle between the instantaneous surface normal at $\bs_i^*$ and one of the two O-H bond vectors of the $i$th water molecule, and  $\theta_i^{(2)}$ is similarly defined for the other O-H bond vector of that same molecule.  This distribution at several values of $x$, where proximity $a_i$ is defined with reference to the instantaneous interface, is shown in \ref{fig:joint_prob}. 

The panels of that figure show how water molecules near the instantaneous liquid-vapor interface adopt orientations consistent with locally favorable hydrogen bond patterns.  In the first layer of liquid, where $n(x)$ has its first peak ($x \approx 1.7 $\AA), the figure shows that water molecules most likely adopt orientations where both OH bonds are aligned to donate hydrogen bonds to other water molecules residing in that first layer.  The geometry of a water molecule makes it impossible for a molecule in the first layer to donate in two such hydrogen bonds when $x \lesssim 0.8$\AA and $x \gtrsim 2.6$\AA. Due to this constraint, the population of water molecules at $x\approx 0.8$\AA most likely orient only one O-H  to donate hydrogen bond to the first layer of liquid while the other OH bond pointed into the vapor phase ($\ell \approx 1$). Similarly, water molecules near $x \approx 2.6$\AA most likely donate one hydrogen bond into the first layer and another into the second layer.  

While not illustrated explicitly here, we find that orientational structure extends from the instantaneous interface into the bulk liquid as far as $8$\AA. For $x \gtrsim 8 $\AA, the solvent orientational structure ceases to exhibit the influence of the interface.

\section*{Methods}
The numerical simulation study consisted of SPC/E water molecules~\cite{SPCE} in a $36\times36\times100$\AA$^3$ simulation cell periodically replicated in the $x$ and $y$ Cartesian directions. At a temperature of 298 K the 1387 water molecules in the simulation cell form a liquid slab spanning the periodic boundaries that is approximately $36$\AA~thick (in the $z$ Cartesian direction). The liquid-vapor phase boundaries present in the simulation serve as a natural barostat and therefore the liquid can be regarded as a system being held at constant pressure. Electrostatic interactions are treated with two-dimensional Ewald summation~\cite{SD79}, and molecular constraints are enforced with the RATTLE algorithm~\cite{HA83}. Statistics were generated through six independently equilibrated  molecular dynamics simulations each run with a time step of 2.4fs for about 1ns with nuclear coordinates written out every 50 time steps.

To identify the water-vapor interface the density field $\bar{\rho}(\mathbf{r},t)$ was computed on a cubic lattice with a lattice spacing of $1 $\AA. For the spatial coarse-graining of the density field, $\phi(\mathbf{r};\xi)$ was truncated and shifted to be both continuous and zero at a distance of $3\xi$. The Gaussian width $\xi$ was selected by first computing a measure of the average amount of interfacial area in the system,
\begin{equation}
A = \left \langle \frac{1}{L^2}\int \Theta[\bar{\rho}(\mathbf{r},t) - c] \Theta [c - \bar{\rho}(\mathbf{r}+\l\hat{z},t)] d\mathbf{r} \right \rangle,
\end{equation}
where $\Theta(x)$ is the Heaviside function which is given by,
$$
\Theta(x) = \left\{ \begin{array}{ll}
		1 , & \quad \mbox{if  $x \geq 0$}, \\
        	0, & \quad \mbox{if $x < 0$.} \end{array} \right.
$$
We have found that a value of $\l = 1$\AA is sufficiently small to ensure an accurate measurement of $A$. For small values of $\xi$ the quantity $A$ decreases with increasing $\xi$, eventually reaching a constant value of $A = L^2$ the projected area of a single planar interface. For large values of $\xi$, $A = L^2$, the projected area of a single planar interface. For smaller values of $\xi$, $A > L^2$, which happens when the density field $\bar{\rho}(\mathbf{r},t)$ in the liquid contains cavities and/or the planar interface develops overhangs. We find that a value of $\xi = 2.4$\AA is large enough to essentially eliminate the occurrence of interface overhangs and bubbles within the liquid phase. Computing the entire density field in this manner for a single configuration (time step) took 3.4 seconds on a single processor of a modern desktop computer.

\section*{Acknowledgements}
Thanks to Kafui Tay and Patrick Varilly for sharing their results rendered for the melittin interface, and Phillip Geissler and John Weeks for useful discussions. This research has been supported by the Director, Office of Science, Office of Basic Energy Sciences, Chemical Sciences, Geosciences, and Biosciences Division, U.S. Department of Energy under Contract No. DE-AC02-05CH11231(APW), and the National Institute of Health through grant R01 GM079102 (DC).

\bibliographystyle{unsrt}

\begin{mcitethebibliography}{31}
\bibitem[Gib()]{Gibbs}
 The Gibbs dividing surface, $\bs_{\mathrm{G}}$, at liquid-vapor coexistence is
  the solution to $\langle \rho(\bs_{\mathrm{G}}) \rangle = c$~\cite{Widom}.
  Provided the mean density profile does not vary quickly in space, and
  provided the system is small enough that $\langle | \bs - \langle \bs \rangle
  | ^2 \rangle$ is not too large, one may show by Taylor expansion that
  $\langle \bs \rangle \approx \bs_{\mathrm{G}}$.\relax
\mciteBstWouldAddEndPunctfalse
\mciteSetBstMidEndSepPunct{\mcitedefaultmidpunct}
{}{\mcitedefaultseppunct}\relax
\EndOfBibitem
\end{mcitethebibliography}

\providecommand*{\mcitethebibliography}{\thebibliography}
\csname @ifundefined\endcsname{endmcitethebibliography}
{\let\endmcitethebibliography\endthebibliography}{}

\end{document}